\documentclass[11pt]{article}
\usepackage{moriond}

\usepackage{relsize}
\def\babar{\mbox{\slshape B\kern-0.1em{\smaller A}\kern-0.1em
    B\kern-0.1em{\smaller A\kern-0.2em R}}}

\def\be{\begin{equation}}
\def\ee{\end{equation}}
\def\bea{\begin{eqnarray}}
\def\eea{\end{eqnarray}}

\newcommand{\Photo}{}

\begin{document}
\vspace*{4cm}
\title{LOW MASS NEW PHYSICS SEARCH FOR A CP-ODD HIGGS BOSON $A^0$ DECAYING TO $s\bar s$ OR GLUON GLUON AT $\babar$}

\author{ E. GUIDO\\
on behalf of the $\babar$ Collaboration }

\address{Universit\`a degli Studi di Genova, Department of Physics, Via Dodecaneso 33 , I-16146 Genova, Italy}

\maketitle\abstracts{
We report on the search for the decay $\Upsilon(1S)\to\gamma A^0$, $A^0 \to gg$ or $s\bar s$, where $A^0$ is the pseudoscalar light Higgs boson predicted by the next-to-minimal supersymmetric standard model. A sample of $\sim18\times 10^6$ $\Upsilon(1S)$ resonances, produced in the $\babar$ experiment via $e^+e^- \to \Upsilon(2S) \to \pi^+\pi^-\Upsilon(1S)$, is used for this search. No significant signal has been found, and upper limits at the $90\%$ of confidence level are set on the product branching fraction of the process.}

\section{Introduction}

The Next to Minimal Supersymmetric Standard Model (NMSSM)~\cite{ref1}, one of the several extensions of the Standard Model, predicts a larger Higgs sector, with two charged, three neutral CP-even, and two neutral CP-odd Higgs bosons. In particular, the model includes the possibility that one of the pseudoscalar Higgs bosons, denoted as $A^0$ hereafter, can be lighter  than two bottom quarks~\cite{ref2}, therefore making its production accessible at the B-factories, via the radiative decay of an $\Upsilon$ resonance.

The $A^0$ is a superposition of a singlet and a non-singlet state, and the value of the branching fraction of the radiative decay $\Upsilon\to\gamma A^0$ actually depends on the non-singlet fraction. The final state to which the $A^0$ decays depends instead on various parameters, such as $\tan\beta$ and the mass of the CP-odd Higgs boson itself~\cite{ref3}. In order to be sensitive to as much parameter space as possible, $\babar$ has performed searches for different final states: $A^0$ decaying into $\mu^+\mu^-$~\cite{ref4,ref5}, into $\tau^+\tau^-$~\cite{ref6,ref7}, into invisible states~\cite{ref8}, and into hadrons~\cite{ref9}, without seeing any significant signal.

The search presented here~\cite{ref10} focuses on the decays $A^0\to gg$ or $s\bar s$. For an $A^0$ mass smaller than $2m_\tau$ , the light pseudoscalar Higgs boson is predicted to decay mostly into two gluons if $\tan\beta$ is of order 1, and into $s\bar s$ if $\tan\beta$ is of order 10. Despite being motivated by NMSSM, the results of this search can be applied to any CP-odd hadronic resonances produced in the radiative decays of $\Upsilon(1S)$.

\section{Experimental technique}

This analysis uses the data recorded by the $\babar$ detector at the PEP-II asymmetric-energy $e^+e^-$ collider at the SLAC National Accelerator Laboratory. The $\babar$ detector is described in detail elsewhere~\cite{ref11,ref12}. We use $\sim$14 fb$^{-1}$ of data taken at the $\Upsilon(2S)$ resonance. Tagging the dipion in the $\Upsilon(2S) \to \pi^+\pi^-\Upsilon(1S)$ transition allows to significantly reduce the otherwise dominant $e^+e^- \to q\bar q$ background, where $q$ is a $u$, $d$, or $s$ quark. We also use $\sim$ 1.4 fb$^{-1}$ of data taken 30~MeV below the $\Upsilon(2S)$ resonance as a background estimate. Simulated signal events with various $A^0$ masses ranging from 0.5 to 9.0 GeV/c$^2$ are used in this analysis.

The final states analyzed must contain: two charged tracks as the dipion system candidate, a radiative photon with an energy greater than 200~MeV when calculated in its center-of-mass frame, and a hadronic system. An exclusive reconstruction of $A^0 \to gg$ is performed, using 26 channels as listed in Table~\ref{tab:tab_01}, while disregarding two-body decay channels because a CP-odd Higgs boson cannot decay into two pseudoscalar mesons. The $A^0 \to s\bar s$ sample is defined as the subset of the 26 $A^0 \to gg$ decay channels that include two or four kaons (channels 11-24 in Table~\ref{tab:tab_01}). Charged kaons, pions, and protons are required to be positively identified.

\begin{table}[t]
\caption{Decay modes for candidate $A^0 \to gg$ and $s\bar s$ decays, sorted by the total mass of the decay products.}
\label{tab:tab_01}
\vspace{0.4cm}
\begin{center}
\begin{tabular}{ll|ll}
\hline
$\#$ & Channel & $\#$ & Channel \\
\hline
1        & $\pi^+\pi^-\pi^0$   & 14 & $K^+K^-\pi^+\pi^-$\\
2        & $\pi^+\pi^-2\pi^0$ & 15 & $K^+K^-\pi^+\pi^-\pi^0$\\
3        & $2\pi^+2\pi^-$      & 16 & $K^\pm K_S^0\pi^\mp\pi^+\pi^-$\\
4        & $2\pi^+2\pi^-\pi^0$ & 17 & $K^+K^-\eta$\\
5        & $\pi^+\pi^-\eta$ & 18 & $K^+K^-2\pi^+2\pi^-$\\
6        & $2\pi^+2\pi^-2\pi^0$ & 19 & $K^\pm K_S^0\pi^\mp\pi^+\pi^-2\pi^0$\\
7        & $3\pi^+3\pi^-$ & 20 & $K^+K^-2\pi^+2\pi^-\pi^0$\\
8        & $2\pi^+2\pi^-\eta$ & 21 & $K^+K^-2\pi^+2\pi^-2\pi^0$\\
9        & $3\pi^+3\pi^-2\pi^0$ & 22 & $K^\pm K_S^0\pi^\mp2\pi^+2\pi^-\pi^0$\\
10      & $4\pi^+4\pi^-$ & 23 & $K^+K^-3\pi^+3\pi^-$\\
11      & $K^+K^-\pi^0$ & 24 & $2K^+2K^-$\\
12      & $K^\pm K_S^0\pi^\mp$ & 25 & $p\bar p \pi^0$\\
13      & $K^+K^-2\pi^0$ & 26 & $p\bar p \pi^+\pi^-$\\
\hline
\end{tabular}
\end{center}
\end{table}

The $A^0$ mass resolution is improved by constraining the $A^0$ candidate and the photon to have an invariant mass equal to the $\Upsilon(1S)$ one, and a decay vertex at the beam spot. 
The main backgrounds to this search are:
\begin{itemize}
\item $\Upsilon(1S)\to \gamma gg$ events, with gluons hadronizing to more than one daughter; it is dominant at low masses, $i.e.$ between 2 and 4 GeV/c$^2$;
\item $\Upsilon(1S)\to ggg$ events, with a $\pi^0$ mistaken as a photon; it is dominant at higher masses, $i.e.$ between 7 and 9 GeV/c$^2$.
\end{itemize}
This search relies on the hadronization modelling used in simulations; the agreement between data and Monte Carlo samples is checked on $\Upsilon(1S)\to \gamma gg$ events, resulting in a scaling factor and a global systematic uncertainty of $50\%$ to be applied to the efficiency. This is the dominant contribution to the systematic uncertainties of this analysis.

\section{Results}

\begin{figure}
\centerline{\includegraphics[width=0.8\linewidth]{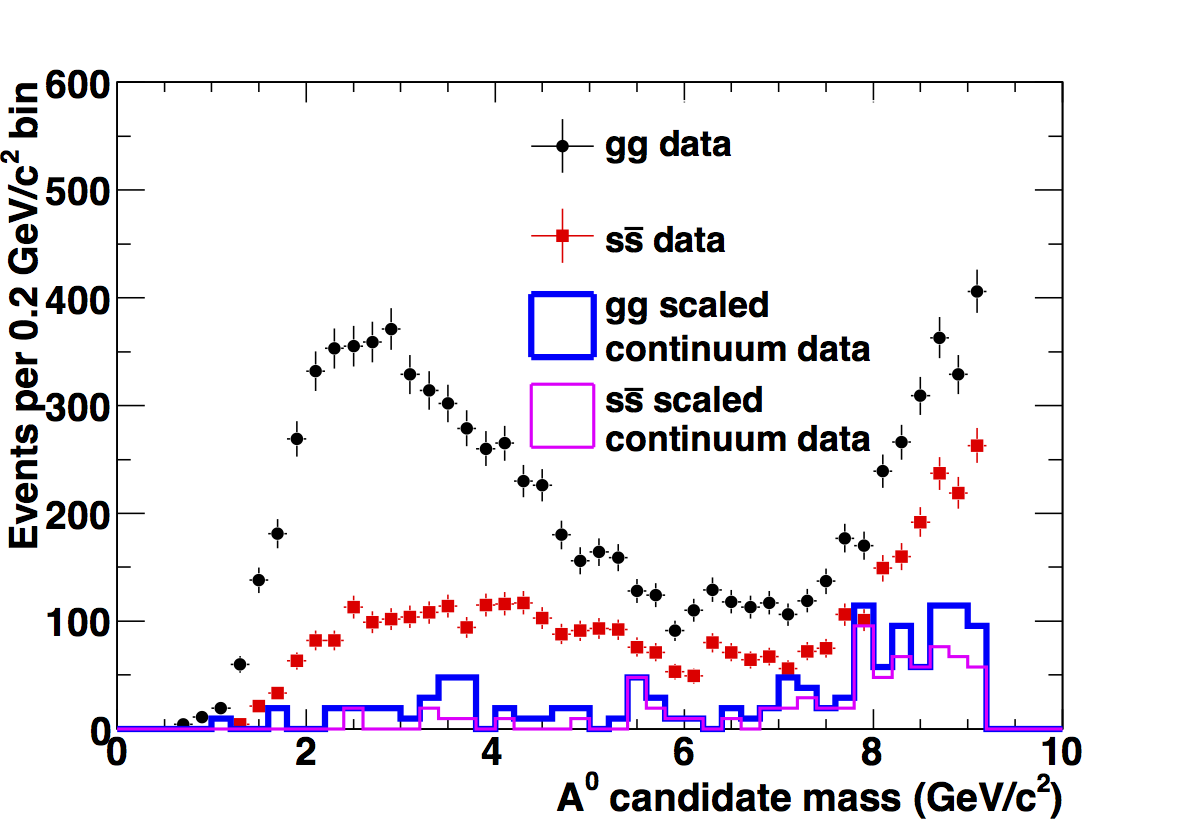}}
\caption{$A^0$ candidate mass spectra after applying all selection criteria. We reconstruct $A^0 \to gg$ using the 26 channels listed in Table~\ref{tab:tab_01} and $A^0\to s\bar s$ using the subset of the same 26 channels that includes two or four kaons. The $A^0$ candidate mass is the invariant mass of the reconstructed hadrons in each channel. The black points with error bars are on-resonance data for $A^0 \to gg$. The red squares with error bars are on-resonance data for $A^0 \to s\bar s$. The thick blue histogram is $A^0 \to gg$ in off-resonance data normalized to the on-resonance integrated luminosity. The thin magenta histogram is $A^0 \to s\bar s$ in off-resonance data normalized to the on-resonance integrated luminosity.}
\label{fig:fig_01}
\end{figure}

\begin{figure}
\centerline{\includegraphics[width=0.8\linewidth]{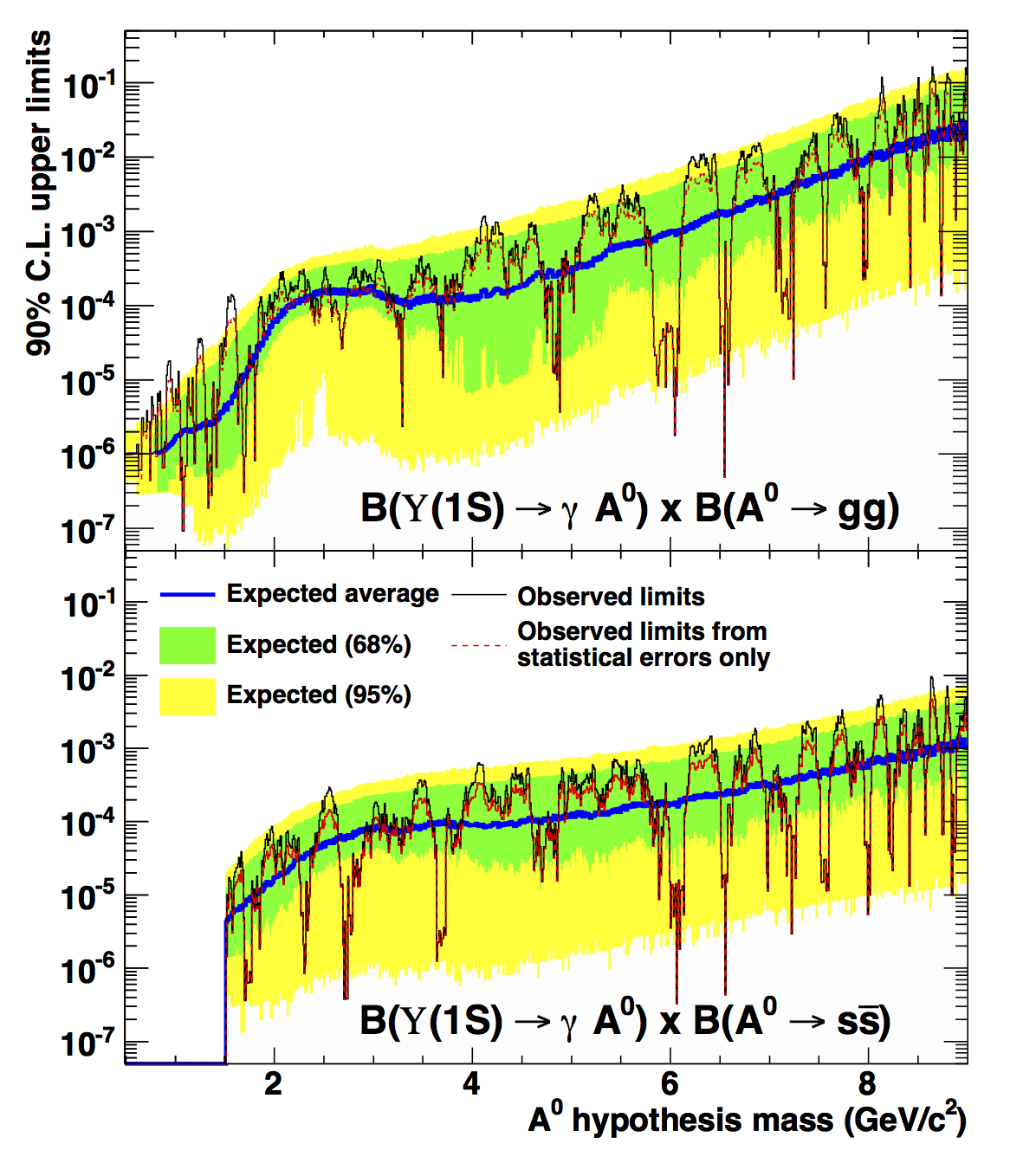}}
\caption{The 90$\%$ confidence level upper limits (thin solid line) on the product branching fractions $\cal{B}$($\Upsilon (1S) \to \gamma A^0)\times$ $\cal{B}$($A^0 \to gg$) (top) and $\cal{B}$($\Upsilon (1S) \to \gamma A^0)\times$ $\cal{B}$($A^0 \to s\bar s$) (bottom). We overlay limits calculated using statistical uncertainties only (thin dashed line). The inner band is the expected region of upper limits in 68$\%$ of simulated experiments. The inner band plus the outer band is the expected region of upper limits in 95$\%$ of simulated experiments. The bands are calculated using all uncertainties. The thick line in the center of the inner band is the expected upper limits calculated using simulated experiments.}
\label{fig:fig_02}
\end{figure}

The candidate mass spectrum is shown in Fig.~\ref{fig:fig_01}. The $A^0$ would appear as a narrow peak in the distribution. A scan of the mass spectrum has been performed in 10 MeV/c$^2$-steps, from 0.5 to 9 GeV/c$^2$, without finding any significant signal through the entire mass range analyzed. Bayesian upper limits at the $90\%$ of confidence level are then set on the product of branching fractions $\cal{B}$($\Upsilon (1S) \to \gamma A^0)\times$ $\cal{B}$($A^0 \to gg$) and $\cal{B}$($\Upsilon (1S) \to \gamma A^0)\times$ $\cal{B}$($A^0 \to s\bar s$), ranging between $10^{-6}$ and $10^{-2}$, and between $10^{-5}$ and $10^{-3}$ for the two final states, respectively, as shown in Fig.~\ref{fig:fig_02}. As a result, the low mass region for $A^0$ is excluded, and no evidence either for a light pseudoscalar Higgs boson, or for any narrow hadronic resonance is found through the entire mass spectrum.

\end{document}